\documentclass[12pt,english,preprint]{emulateapj}
\usepackage{graphicx}
\usepackage{amssymb}

\makeatletter

\providecommand{\tabularnewline}{\\}

\usepackage{array}

\makeatletter

\usepackage{times}

\shorttitle{Binary rejuvenation and triple disruption}
\shortauthors{Perets}

\makeatother

\usepackage{babel}
\makeatother

\begin{document}
\newcommand{\Ms}{M_{\star}} \newcommand{\Rs}{R_{\star}} \newcommand{\Ls}{L_{\star}}

\newcommand{\Mo}{M_{\odot}} \newcommand{\Ro}{R_{\odot}} \newcommand{\Lo}{L_{\odot}}
\newcommand{\Mbh}{M_{\bullet}} \newcommand{\np}{n_{p}} \newcommand{\Np}{N_{P}}
\newcommand{\Mp}{M_{p}} \newcommand{\Ns}{N_{\star}} \newcommand{\SgrA}{SgrA^{\star}}
\newcommand{\rMP}{r_{\mathrm{MP}}}


\title{Runaway and hypervelocity stars in the Galactic halo:\\
Binary rejuvenation and triple disruption }

\author{Hagai B. Perets\altaffilmark{1} }

\email{hagai.perets@weizmann.ac.il}

\affil{Weizmann Institute of Science, POB 26, Rehovot 76100, Israel}

\begin{abstract}
Young stars observed in the distant Galactic halo are usually thought
to have formed elsewhere, either in the Galactic disk or perhaps the
Galactic center, and subsequently ejected at high velocities to their
current position. However, some of these stars have apparent lifetimes
shorter the required flight time from the Galactic disk/center. We
suggest that such stars have evolved in close runaway or hypervelocity
binaries. Stellar evolution of such binaries can drive them into mass
transfer configurations and even mergers. Such evolution could then
rejuvenate them (e.g. blue stragglers) and extend their lifetime after
their ejection. The extended lifetimes of such stars could then be
reconciled with their flight times to the Galactic halo. We study
the possibilities of binary runaway and hypervelocity stars and show
that such binaries could have been ejected in triple disruptions and
other dynamical interactions with stars or with massive black holes.
We show that currently observed {}``too young'' star in the halo
could have been ejected from the Galactic disk or the Galactic center
and be observable in their current position if they were ejected as
binaries. Specifically it is shown that the hypervelocity star HE
0437-5439 could be such a rejuvenated star. Other suggestions for
its ejection from the LMC are found to be highly unlikely. Moreover, 
it is shown that its observed metallicity is most consistent with a 
Galactic origin and a Galactic center origin can not currently rule out.
In addition, we suggest that triple disruptions by the massive black
hole in the Galactic center could also capture binaries in close orbits
near the MBH, some of which may later evolve to become more massive
rejuvenated stars. 
\end{abstract}


\section{Introduction}

Young massive OB stars are usually observed close to their birth place
in young stellar clusters or associations \citep[e.g. ][]{hoo+00}.
Some of them, however, are observed in isolation, far from any star
forming region. Observations of such stars in the Galactic halo are
especially puzzling given the unfavorable conditions for regular star
formation in such regions. Formation of massive stars usually requires
special conditions such as the the existence of molecular clouds with
dense cores that could collapse to form massive stars. Such gas reservoirs
do not exist at large distances in the Galactic halo \citep{sav+81,sem+94}.
Nevertheless, young OB stars are observed there. These halo young
stars are usually observed to have high peculiar velocities ($>40\, km\, s^{-1}$),
and are thought to have been ejected from their birth place and acquire
their high velocity through some dynamical process. Due to their high
velocities they could propagate to their currently observed remote
location, even during their short lifetimes. Such high velocity stars
are thought to be dynamically ejected in stellar binary interactions
or through a binary supernova explosions (so called runaway stars;
\citealt[e.g. ][and references there in]{bla61,pov+67,mar06}). Young
stars with even higher velocities, so called Hypervelocity stars (HVSs;
with velocities $>300\, km\, s^{-1}$; \citealp[and references there in]{hil88,bro+05,hir+05,bro+07b}),
are most likely ejected from the Galactic center (GC) through some
interaction with the massive black hole (MBH) known to exist there
(\citealp{hil88,yuq+03}). However, in some cases even the high velocities
of these runaway/hypervelocity stars are not sufficient to explain
their remote locations, so far from any star forming region. In those
cases it appears that flight times of these the young stars from their
birthplace in the Galactic disk (or the GC) to their current location
is longer than their main sequence (MS) lifetimes. 

In the following we suggest that such discrepancy could be solved
if these stars were ejected as runaway or hypervelocity binaries.
A combination of dynamical and evolutionary processes could then explain
the existence of these {}``too young'' halo stars. Following their
ejection and propagation in the Galaxy, the ejected runaway/hypervelocity
binaries could evolve and rejuvenate through mass between (or merger
of) the binary stellar components. Such rejuvenation could extend
the main sequence lifetime of these stars and resolve the discrepancy
between their apparent lifetimes and estimated flight times. We first
shortly overview the observations of Galactic halo young stars in
\S2. We then discuss the possible dynamical scenarios for the ejection
of runaway and hypervelocity binaries that can serve as progenitors
of rejuvenated halo stars (\S3). The rejuvenation scenario is presented
in \S4, followed by the discussion (\S5) and summary.

\section{Young stars in the Galactic halo}

\label{sec:halo young stars}

In many cases dynamical processes can eject massive stars from their
original birth place at high velocities. These so called 'runaway'
stars (\citealp{bla61}; for a short overview see \citealp{hoo+01})
constitute a considerable fraction of the early O and B star population
in the galaxy;$\sim30-40\%$ of the O stars and $5-10\%$ of the B
stars (\citealp{sto91}, and refs. within). Such stars have large
peculiar velocities of $40\le v_{pec}\le200\, km\, s^{-1}$ \citep{gie87,hoo+01}
or even higher \citep{mar06}. Besides their relatively high velocities,
runaway stars are also distinguished from the normal early-type stars
by their much lower ($<10\%$) multiplicity compared with the binary
fraction of normal early-type stars ($>50\%$ and up to $100\%$;
\citealt{gar+80,mas+98,kob+07,kou+07}). An additional class of stars
with even higher velocities have been discovered in recent years,
with the observations of several HVSs ($>300\, km\, s^{-1}$) in the
Galactic halo \citep{bro+05,ede+06,bro+07b}.

In most cases observed OB runaways and HVSs are found to have kinematics
consistent with an ejection from a star forming region in the Galactic
disk of from the GC \citep{hoo+01,mar06,bro+07b}. However, some of
the young stars observed at large distances in the Galactic halo can
not be regular runaway or hypervelocity stars. The flight times of
these stars, required in order to reach their current position after
being ejected from their birth place, are longer than their lifetimes
(given their observed positions and velocities). Table \ref{t:young-halo-stars}
shows a list of such candidate stars, collected from the literature,
where we list stars with calculated propagation times which could
potentially be larger than their evolutionary time, given the uncertainties.
We caution that the uncertainties in the radial and/or proper motion
velocities exist for some of these stars, and their travel time and
evolutionary times may be in fact consistent (this is true for 10
of the 16 stars in the table). Nevertheless, we show all of the halo
stars that potentially have travel time and life time inconsistencies,
not excluded by current observations. Several other stars suggested
in the literature to have such inconsistencies, were later found to
be closer/older or faster stars in later observations (\citealt{ram+01,lyn+04,mar06}).
Such stars were excluded from the list. 

Such 'too young' halo stars have been suggested to possibly form in
situ in the halo (see \citealt{kee+92} and references there in).
However, as we suggest below, it is far more likely that these stars
have been ejected as binaries and then became blue stragglers by rejuvenating
through mass transfer from (or merger with) their binary companion. 

\begin{table}
\caption{\label{t:young-halo-stars}Too young stars in the Galactic halo}

\begin{centering}
\begin{tabular}{lcccl}
\hline 
Name  & Mass & $T_{evol}$  & $T_{ej}$   & References\tabularnewline
 & ($M_{\odot}$) & (Myrs) & (Myrs) & \tabularnewline
\hline 
\multicolumn{5}{l}{Hypervelocity Stars}\tabularnewline
\noalign{\vskip\doublerulesep}
\hline 
$\,\,$HE 0437-5439 & $9$ & $29$ & $90\pm10$ & \citealt{ede+06}\tabularnewline
\noalign{\vskip\doublerulesep}
\hline 
Runaway stars &  &  &  & \tabularnewline
\hline
\noalign{\vskip\doublerulesep}
$\,\,$BD +38 2182 & $6.6$ & $60$ & $53\pm8$ & \citealt{mar06}\tabularnewline
$\,\,$BD +36 2268 & $6.9$ & $51$ & $50\pm2$ & {}``\tabularnewline
$\,\,$HD 140543 & $22$ & $8$ & $23\pm3$ & {}``\tabularnewline
$\,\,$HD 188618 & $9.7$ & $26$ & $38\pm8$ & {}``\tabularnewline
$\,\,$HD 206144 & $9.7$ & $26$ & $25\pm2$ & {}``\tabularnewline
$\,\,$PG 0122+214 & $6.7$ & $35\pm6$ & $51\pm24$ & \citealt{ram+01}\tabularnewline
$\,\,$PG 1610+239 & $5.8$ & $54\pm10$ & $>62$ & {}``\tabularnewline
$\,\,$PHL 159 & $8$ & $28\pm2$ & $31$ & {}``\tabularnewline
$\,\,$PHL 346 & $9.9$ & $19\pm2$ & $27\pm7$ & {}``\tabularnewline
$\,\,$SB 357 & $7.4$ & $26\pm4$ & $61$ & {}``\tabularnewline
$\,\,$HS 1914+7139 & $6.2$ & $39\pm6$ & $91$ & {}``\tabularnewline
$\,\,$PG 0914+001$^{a}$ & $5.8\,(4.7)$ & $79\,(116)$ & $199\,(109)$ & \citealt{lyn+04}\tabularnewline
$\,\,$PG 1209+263$^{a}$ & $6.3\,(5)$ & $62\,(91)$ & $272\,(170)$ & {}``\tabularnewline
$\,\,$PG 2219+094$^{a}$ & $7.5\,(6.5)$ & $41\,(53)$ & $53\,(52)$ & {}``\tabularnewline
$\,\,$PG 2229+099$^{a}$ & $5.8\,(5.4)$ & $49\,(63)$ & $63\,(58)$ & {}``\tabularnewline
\hline 
\multicolumn{5}{l}{$^{a}${\footnotesize Numbers in parentheses show lower/upper limits
on the timescales }}\tabularnewline
\multicolumn{5}{l}{{\footnotesize $\,\,$that could minimize the time discrepancies (from
\citealt{lyn+04}).}}\tabularnewline
\hline
\end{tabular}
\par\end{centering}
\end{table}

\section{Dynamical ejection of runaway and hypervelocity binaries}

\subsection{Runaway binaries}

Two mechanisms are thought to contribute to the ejection of runaway
stars, both involve binarity (or higher multiplicity). In the binary
supernova scenario (\citealt{bla61}) a runaway star receives its
velocity when the primary component of a massive binary system explodes
as a supernova (SN). When the SN shell passes the secondary, the gravitational
attraction of the primary reduces considerably, and the secondary
starts to move through space with a velocity comparable to its original
orbital velocity. In the dynamical ejection scenario (\citealt{pov+67})
runaway stars are formed through gravitational interactions between
stars in dense, compact clusters. Simulations show that such encounters
may produce runaways with velocities up to $200\, km\, s^{-1}$ (\citealt{mik83,leo+90,leo+91,gua+04}).
These scenarios suggest that many of the early OB stars formed in
young clusters could be ejected from their birth place and leave the
cluster at high velocity. 

Theoretical studies suggest that binary stars could also be ejected
at high velocities, although at smaller fraction of $\sim0.1$ of
all the runaway stars \citep{leo+88,leo+90,por00}. Such runaway binaries
have indeed been observed \citep{gie+86,mas+98,mar06,loc+07b,mcs+07a,mcs+07c},
with fractions of \textasciitilde{}0.1 in the runaway stars samples.
The periods of the runaway binaries were found to be typically short
($<5$ days; \citealt{gie+86,mas+98,mar03}) as expected from the
dynamical ejection scenario. Some of the binaries were found to be
with larger period (\textasciitilde{}20 days) and relatively eccentric
orbits ($>0.4$) and are thought to be ejected due to a SN explosion
\citep{loc+07b,mcs+07a,mcs+07c}, in which case rejuvenation is not
possible.

\subsection{Hypervelocity binaries}

Extreme velocities as found for HVSs most likely suggest a different
dynamical origin than that of runaway stars. Several scenarios have
been suggested for ejection of HVSs, all of them require an interaction
with the MBH. These include a disruption of a stellar binary by a
MBH (\citealt{hil88,yuq+03,gin+06,per+07}), an interaction of a single
star with an intermediate mass black hole (IMBH) which inspirals to
the GC (\citealt{han+03a,yuq+03,lev05,bau+06,loc+07a,ses+07b}), or
interaction with stellar black holes (SBHs) in the GC (\citealt{yuq+03,mir+00,ole+07}).
In such scenarios stars could be ejected from the GC with velocities
of hundreds and even a few thousands $km\, s^{-1}$ possibly extending
much beyond the escape velocity from the galaxy. 

Recently it was suggested that binary stars could also be ejected
as hypervelocity binaries during the inspiral of an IMBH \citep{luy+07,ses+08},
and could serve as evidence for the binary MBH ejection scenario.
It was noted that the other scenarios for hypervelocity ejection are
not likely to eject hypervelocity binaries. Specifically the probability
for a binary ejection in a triple disruption by a MBH is negligibly
small. The later claim may well be correct for the low mass hypervelocity
stars discussed in Lu et al. paper, however, as we show in the following,
a non-negligible number of massive hypervelocity binaries could be
ejected through a triple disruption by the MBH in the GC. Young massive
binaries may also be ejected by an inspiraling IMBH as suggested by
Lu et al. for low mass binaries. However, the fraction of surviving
binaries close to the MBH, where they could be ejected as HVSs by
an inspiraling IMBH is small, since most would be disrupted through
dynamical interactions with other stars in this hostile environment;
see \citet{per09} for detailed discussion. Moreover, it is likely
that the most if not all of the observed young HVSs in the Galactic
halo were not ejected in such a scenario, given the observational
constrains on the number of young stars observed close to the MBH
in the GC \citep{per09}. In the following we discuss the triple disruption
scenario.

\subsubsection{Triple disruption by a MBH}

A close pass of a binary star near a massive black hole results in
an exchange interaction, in which one star is ejected at high velocity,
while its companion is captured by the MBH and is left bound to it.
Such interaction occurs because of the tidal forces exerted by the
MBH on the binary components. Typically, a binary (with mass, $M_{bin}=M_{e}+M_{c}$
and semi-major axis, $a_{bin}$), is disrupted when it crosses the
tidal radius of the MBH (with mass $M_{BH}$), given by

\begin{equation}
r_{t}=\left(\frac{M_{BH}}{M_{bin}}\right)^{1/3}a_{bin}\label{eq:rtb}\end{equation}
and one of the stars (with mass $M_{c}$) is captured close to the
MBH and the other (with mass $M_{e}$) is ejected at high velocity
of about \citep{hil91,bro+06c} 

\begin{eqnarray}
v_{\mathrm{BH}} & = & 1800\,\mathrm{km\, s^{-1}}\times\nonumber \\
 &  & \left(\frac{a_{bin}}{0.1\,\mathrm{AU}}\right)^{-1/2}\left(\frac{M_{e}+M_{c}}{2M_{\odot}}\right)^{1/3}\!\left(\frac{M_{BH}}{4\!\times\!10^{6}\,\Mo}\right)^{1/6}\left(\frac{2M_{c}}{M_{e}+M_{c}}\right)^{(1/2)}\,.\label{eq:v_eject}\end{eqnarray}

The same scenario could be extended to a triple disruption by a MBH.
Triple stars have a stable configuration if the semi major axis of
the outer binary, $a_{o}$ is much larger than the semi major axis
of the inner binary, $a_{i}$(i.e. $a_{i}\ll a_{o}$). In such hierarchical
triples, the outer binary could be disrupted by the MBH while the
inner closer binary is kept bound. In this case a triple disruption
could produce a hypervelocity binary, or alternatively a captured
binary star near the MBH. The ejection velocity in Eq. \ref{eq:v_eject}
is strongly dependent on the semi major axis of the binary (the outer
binary in the triple case) and on the mass of the stellar binary (triple
in this case). Both of these parameters vary by much between the population
of low mass stars and high mass stars. 

For low mass stars ($M_{triple}\sim3\, M_{\odot}$, for equal mass
stars) such as studied by \citet{luy+07}, one requires $a_{bin}=a_{o}\sim0.4\, AU$
for ejection of a hypervelocity binary at $\sim900\, km\, s^{-1}$.
Such close binaries are infrequent (only a few percents of the binary
population; \citealt{duq+91}), and the fraction of low mass triples
with such close outer binaries is negligibly small \citep{tok+06}. 

For higher mass stars such as the observed young B-type hypervelocity
stars in the Galactic halo, $m_{\star}\sim2-4\, M_{\odot}$ corresponding
to a triple mass of $M_{triple}\sim6-12\, M_{\odot}$ (assuming equal
mass stars). In this case even a semi major axis of $a_{o}\sim0.6-1\, AU$
is sufficient for the ejection of a hypervelocity binary. High mass
binary stars are known to have higher binary fraction (probably $f_{bin}>0.8$,
e.g. \citealt{abt+90,mas+98,kob+07}) and different semi-major axis
distribution than low mass stars, with a large fraction of them $(f_{cbin}\sim0.4$)
in close binaries ($a_{bin}<1\, AU$, e.g. \citealt{abt83,mor+91}).
The triple fraction and distribution of massive stars is still uncertain,
but it is strongly suggestive of a high triple fraction among binaries.
\citet{eva+05} find that most if not all of the massive binaries
they observed (in Cepheids) are likely to be triple systems. Some
$f_{triple}\sim0.8$ of the wide visual binaries in stellar associations
are in fact hierarchical triple systems, where typically the more
massive of the binary components is itself a spectroscopic or even
eclipsing binary pair \citep{zin05}. \citet{fek81} compared the
properties of close multiple stars. He finds a fraction $f_{\frac{1}{2}yr}\sim0.2$
of the more massive systems (we choose systems with total mass of
$>6\, M_{\odot}$) to have outer binary periods shorter than half
a year, corresponding to $a_{o}\lesssim1\, AU$, i.e with characteristics
allowing for the ejection of hypervelocity binary, if they were disrupted
by a MBH. The fraction of close triples (such as those in \citealp{fek81})
out of the total triple population is unknown. In the lack of better
estimate we assume this fraction to follow the the fraction of close
binaries%
\footnote{This is a reasonable assumption since the difference in the dynamics
of a star due to the interaction with a companion single mass or a
very close binary (i.e. the inner binary) are very small, especially
when discussing the surviving triples that are hierarchical. Nevertheless,
we emphasize that this is an assumption. Future observations of triple
systems, when available, should be used to produce more accurate and
not assumption dependent estimates of triples distributions. %
}, i.e. we take the fraction of close triples out of the full triple
population to be $f_{ctriple}=f_{cbin}\cdot f_{\frac{1}{2}yr}=0.4\times0.2=0.08$
(note that the fraction of close triples might in fact be higher,
since $f_{cbin}$ is taken for binaries with $a_{bin}<1\, AU$ where
as Fekel's sample also contain triples with wider outer binaries).
Taken together we can estimate the triple fraction of hypervelocity
binary potential progenitors to be $f_{prog}=f_{bin}\cdot f_{triple}\cdot f_{ctriple}\simeq0.05$
(taking a binary fraction of $0.8$ of which $0.8$ are triples, and
$0.08$ of those have outer binaries with period $<0.5$ yr). 
We note that there is some weak trend for more massive stars 
to have higher multiplicity,
however, more observational data is required for a better resolution
of the mass dependence of the multiplicity, and the quoted values are
assumed to represent all main sequence B-stars. 

Note that very few studies on observed massive triples have been done,
and therefore we also try to estimate the appropriate triple fraction
differently, based only on the better known characteristics of massive
binaries, where we follow the method used by \citet{fab+07}. Again,
we assume that the orbital and stellar characteristics of the third
component in a given triple could be chosen from the same distributions
of the binaries$^{1}$. We pick a sample of randomly chosen triple
systems taken from the appropriate binary distributions. For our sample
we choose many triples with initial orbital distributions such that
both their inner and outer binaries are taken from the best fit observed
distributions of \citet{kob+07}. Each given triple has an inner binary
with semi major axis $a_{i}$ and masses $m_{1}$ and $m_{2}$; and
an outer companion with mass $m_{3}$ in an orbit with semi major
axis $a_{o}$. The period is chosen from a distribution of orbital
separations which is flat in log space, $f(\log p)\propto Const$
, corresponding to $f(r)\propto1/r$ (i.e., $\ddot{O}$pik\textquoteright{}s
law). The minimum value for the for the appropriate semi-major axis
is taken to be twice the radii of the binary stars, i.e. separation
of a contact binary which does not immediately merge. The maximal
value is taken to be 1000 AU, following the best fit distribution
found by \citet{kob+07}. The mass ratio, $q$, is chosen from a power
law distribution ($f(q)\propto q^{-0.4}$); we also tried other distributions
suggested in the literature and found only minor effects on the final
calculated fraction. The mass of the tertiary companion, $m_{3}$,
was determined by choosing $q=m_{3}/m_{1}+m_{2}$ from the same mass
ratio distribution. This approach implies that the mass of the third
star was correlated with the mass of the inner binary, but we do not
believe that this correlation has any significant effect on our results.

Two periods and eccentricities were picked in the same manner discussed
in \citeauthor{fab+07}. The smaller (larger) period was assigned
to the inner (outer) orbit. The semi major axes were computed from
these masses and periods assuming non interacting Keplerian orbits.
The mutual inclination distribution of the tertiary is assumed to
be isotropic with respect to the inner binary. After these parameters
were selected, we used the empirical stability criterion used in \citeauthor{fab+07}
(2007; see their Eq. (37) originally formulated in \citealp{mar+01})
to determine whether the system is hierarchical or if it will disrupt
in a small number of dynamical times. If the semi major axes obeyed
this criterion then we accepted the triple as stable, otherwise, we
assumed it disrupted. We then found the fraction of stable triples,
such that their disruption by the MBH in the GC could eject a hypervelocity
binary, i.e. have outer semi-major axis small enough for the ejection
velocity to be high. 

From a large sample of triples ($10^{5})$ we find that about $0.03$
of potentially formed triples are stable triples that could serve
as hypervelocity binaries progenitors, where this results is not very
sensitive to the prime mass $m_{1}$ of the inner binary component.
This is generally consistent with the observation based estimates
given before. We conclude that $\sim0.03-0.05$  of all massive hypervelocity
stars could have been ejected with a binary companion or have left
a close binary captured in an orbit very close to the MBH.

\section{Rejuvenation and evolution in high velocity close binaries}

\label{sec:rej}

Binaries ejected at high velocities are relatively close binaries.
For both runaway and hypervelocity binaries the closer the binary
is, the higher is its probability to be ejected at high velocity (see
e.g. Eq. \ref{eq:v_eject}). In the case of binaries evolving in triples
(e.g. hypervelocity binaries from triple disruptions), dynamical evolution
could be very efficient in producing very close inner binaries. A
large fraction of such triples evolve through Kozai oscillations \citep{koz62},
in which the inner binary is periodically driven into high eccentricities.
When the eccentricity of the inner binaries are high enough the binary
components tidally interact, and dissipate the orbital energy. This
mechanism of Kozai cycles and tidal friction (KCTF) was shown to drive
the inner binary into close configuration and circularization at periods
of a few days \citep{kis+98,egg+01,fab+07}, at relatively short times
($\sim$Myr), much shorter than the MS lifetime of the stars (Fabrycky,
private communication; 2007). In fact it is quite plausible that most
of the observed contact binaries are produced through evolution in
triple stars \citep{dan+06,pri+06a,fab+07}. We conclude that most
if not all runaway and hypervelocity binaries should be ejected as 
close (period of up to a few tens of days) or even contact binaries.

\begin{figure*}
\includegraphics[clip,scale=0.4]{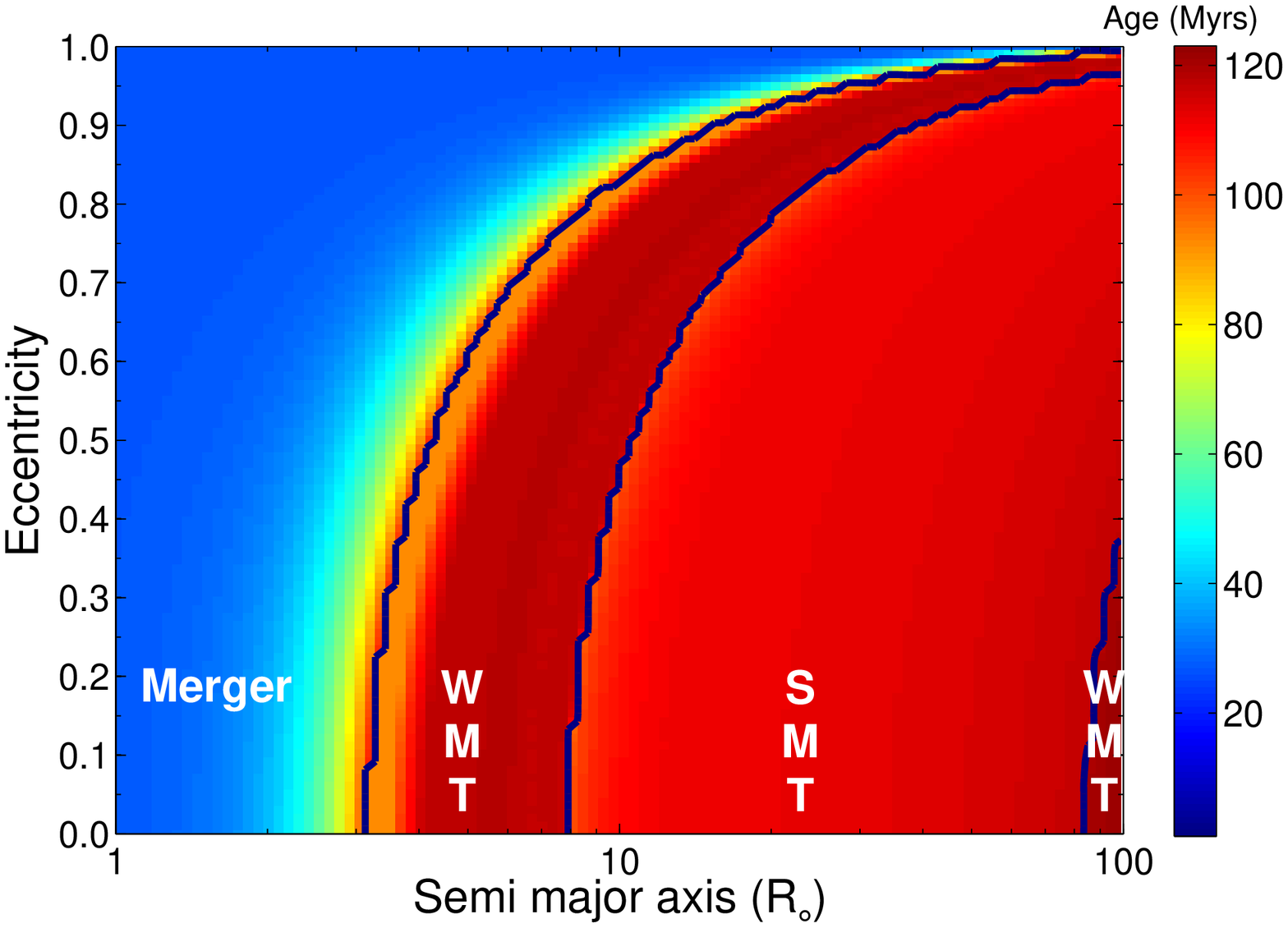}\includegraphics[scale=0.4]{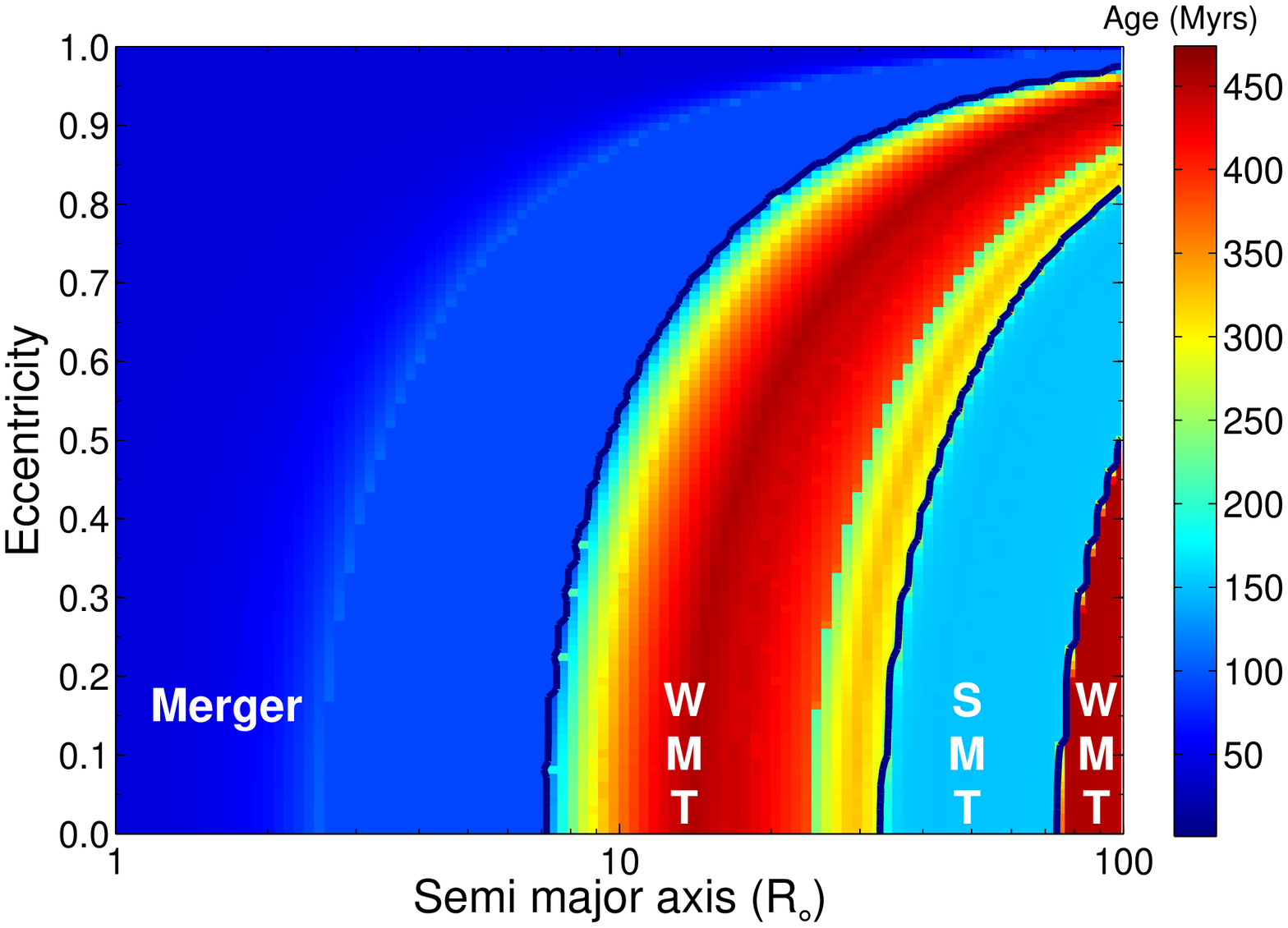}

\includegraphics[scale=0.4]{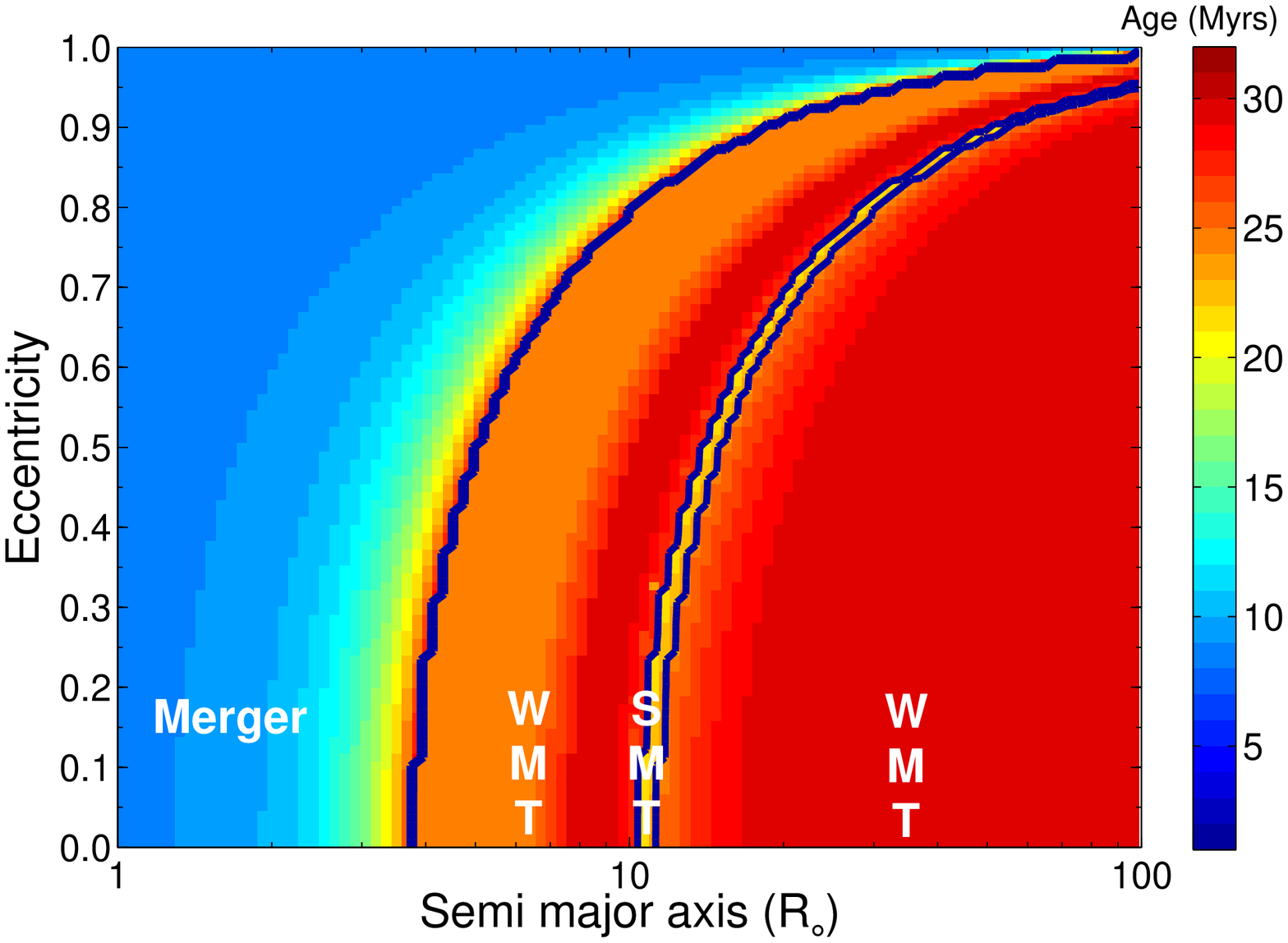}\includegraphics[scale=0.4]{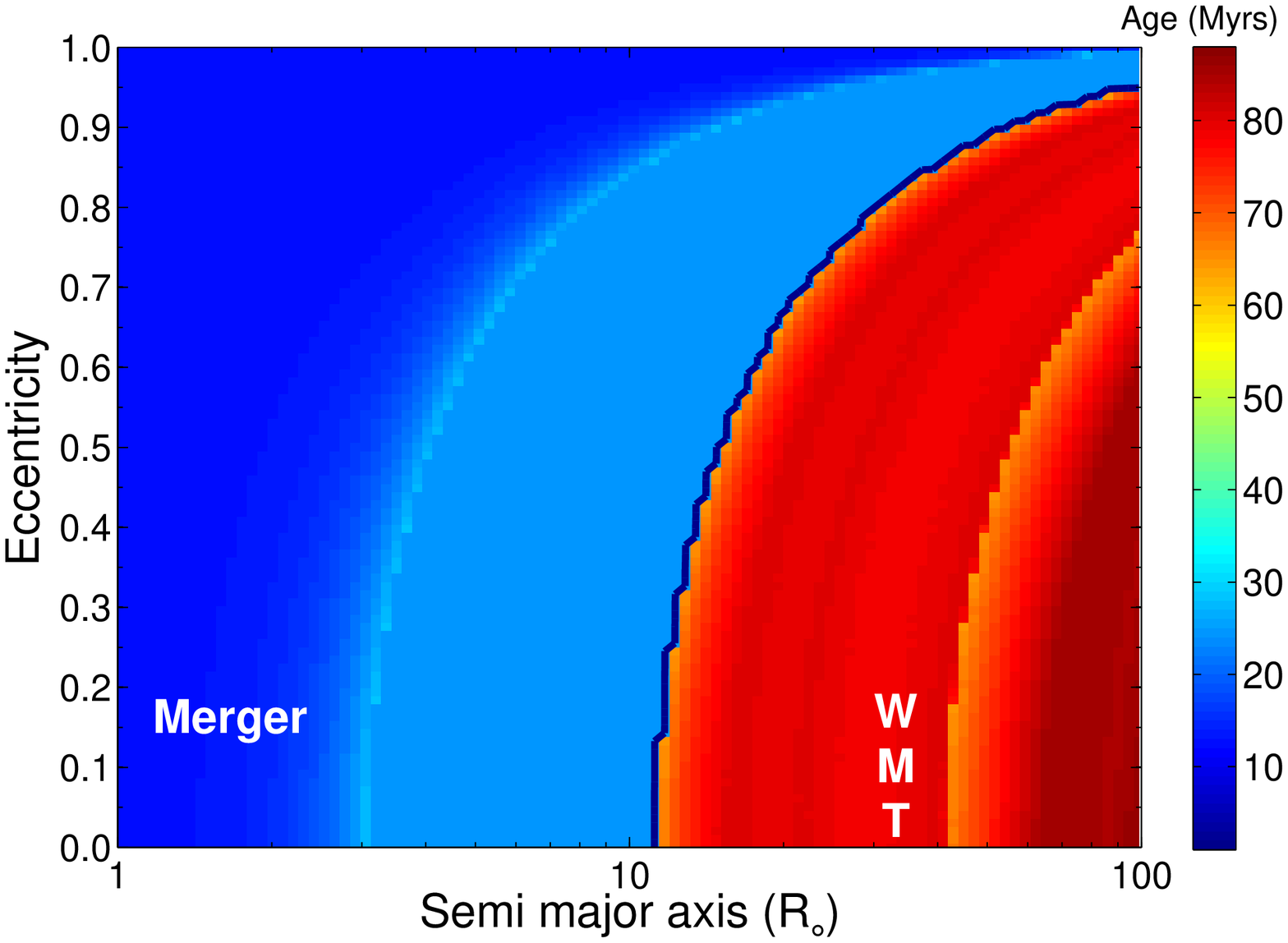}

\caption{\label{fig:Evolutionary-time-scale}Evolutionary time scale for close
binaries with a range of initial conditions (semi-major axis and eccentricity).
Colors (in Myrs) indicate the evolutionary time after which both binary
components end their life on the main sequence. Black lines separate
the different evolutionary regimes; full merger, weak mass transfer
(WMT) and strong mass transfer (SMT). Four cases are shown corresponding
to binaries with prime masses of $4.8\, M_{\odot}$(upper figs.) and
$9.6\, M_{\odot}$(lower figs.), and mass ratios of $0.9$(left figs.)
and $0.5$ (right figs.). The masses of the rejuvenated binary component
are $m_{rej}>0.95$ times the initial total binary mass $m_{bin}$
in the merger case (single star); $0.8\, m_{bin}<m_{rej}<0.9\, m_{bin}$
in the SMT case; and $0.4\, m_{bin}<m_{rej}<0.6\, m_{bin}$in the
WMT case. }

\end{figure*}

During the evolution of close binaries in the triples they can closely interact through
mass transfer and even mergers (cite{per+09}). Such interaction could lead to rejuvenation
of one the binary components (see e.g \citealt{dra+07,van+98b}),
and even to its {}``reincarnation'' as higher mass star. Such rejuvenated
stars, also known as blue stragglers (although this term usually refers
to low mass rejuvenated stars) could appear much younger than their
real age \citep{van+98b}. For high velocity stars such extended lifetimes
could potentially be translated into much larger propagation distances
from their birth place. Consequently, such high velocity massive blue
stragglers could be observed to have flight times longer than their
apparent lifetime (see table 1). 

We used the population synthesis program SeBa%
\footnote{http://www.ids.ias.edu/\textasciitilde{}starlab/seba/%
}, as described in detail in \citet{por+96} and \citet{por+98}, to
study the possible outcomes from the evolution of close runaway and
hypervelocity binaries. In this module stars are evolved via the time
dependent mass-radius relations for solar metallicities given by \citet{egg+89}
with corrections by \citet{egg+90} and \citet{tou+96}. These equations
give the radius of a star as a function of time and the initial mass
of the star (on the zero-age main-sequence). The mass of the stellar
core and the rate of mass loss via a stellar wind (not specified in
this prescription) were included using the prescriptions of \citet{por+96}. 

We have focused on the evolution of the possible binary progenitors
of the rejuvenated OB stars, such as currently observed in the Galactic
halo (table 1). For this purpose we generated a grid of initial conditions
for the periods (in the range $1-100$ days) and eccentricities ($0-0.99$)
of the binaries. Observations of massive binaries show that the components
mass ratios are usually large \citep{kob+07}. We have studied binaries
with different masses and/or mass ratios. For each possibility we
followed the binaries evolution for any initial condition in our grids,
until both the binary components (or single component in case of a
merger) have finished their evolution on the MS. We then recorded
the total evolution time and the binary characteristics (component
masses, period, and eccentricity). Here we show the results of two
representative masses for the prime binary component ($4.8$ and $9.6\, M_{\odot}$)
and two possible mass ratios ($q=0.9,\,0.5$). Similar results have
been found for other masses and mass ratios. 

The binary evolution scenarios in our grid which produce rejuvenated stars
correspond to different types of mass transfer scenarios, namely 
type A and type B mass transfer (\cite{pac71}).   
We find three main outcomes for the evolution of such
close binaries; merger, strong mass transfer and weak mass transfer
(see Fig. \ref{fig:Evolutionary-time-scale}). When the two components
are very close (i.e. small periods of a few days for circular orbits,
or longer period binaries with higher eccentricities), the binary
components merge to form a single star containing almost all of the
total initial binary mass. These mass transfer scenarios are all basically 
subtypes of type A mass transfer (see \cite{nel+01} for detailed 
description).
The time until merger varies between almost
immediate merger up to the MS lifetime of the more massive component.
A mass transfer and/or merger evolution may lead to some observational 
signatures, such as possibly high rotational velocity 
(\citealp[and references there in]{leo95}); or chemical anomalies, such
as CNO abundances anomalies citep{sar+96,che+04,fer+06}. 
For our rejuvenated stars candidate sample we find that high rotational 
velocities are possibly observed in PG 1209+263 , HS 1914+7139; PG 0914+001 \citep{ram+01}.Large chemical anomalies are observed in HS 1914+7139
; \citealt{lyn+04}).
However, such peculiarities are not likely to be strong 
signatures, since they do not necessarily arise only due to mass transfer 
processes, and may not even be produced in the majority of rejuvenated stars
to begin with. We conclude that given the weak observational signature of the 
rejuvenation process on the
appearance of the rejuvenated stars, observations of the chemical or 
rotational properties can not serve to directly trace their binary origin.

In longer period binaries (and/or smaller eccentricities) strong mass
transfer occurs when the more massive binary components leave the
MS. The massive component then shed most of its mass to its companion.
This scenario usually produces a massive MS star containing $0.8-0.9$
of the initial total binary mass with a low mass companion. Such scenario 
corresponds to type B mass transfer (see e.g. \cite{pac71} for details).
 These binaries have typical periods of $few\times10-200$ days. The timescale
for the production of the accreting massive components is typically
the MS lifetime of the initial prime component of the binary. In a
smaller part of the phase space explored in our grids, a weak mass
transfer occurs where only a small fraction of the mass from the prime
component is accreted by its companion. In these cases, the rejuvenation
of the companion is negligible.

\section{Discussion}

Young stars have been observed in the Galactic halo since the 1970's
\citep[e.g. ][]{gre+74}. Such stars are found well away from any
star-forming region and are remote from any high density interstellar
gas pockets where they could have potentially formed. The origin of
these young stars in the Galactic halo have been extensively studied
(see \citealp{kee+92} for a review), and many of them could be understood
in terms of ejection mechanisms. They could have formed in the Galactic
disk and then be ejected at high velocities due to dynamical interactions,
and travel to their current position in the halo. However, some of
these young halo stars have apparent evolutionary age shorter than
the flight time from any star forming region in the Galactic disk
or the GC (cf. table 1). 

In situ star formation in the Galactic halo could potentially explain
the existence of young halo stars. However, the very low gas density
in the halo (e.g. \citealp{sav+81,sem+94}) make this possibility
seem difficult. \citet{dys+83} suggested that star formation could
occur during collisions between cloudlets within high velocity clouds
at high galactic latitudes, but \citet{chr+97} have shown that such
events are much too rare. \citet{mar+99} suggested that spiral density
waves in the disk could trigger star formation above the Galactic
plane up to a kpc, but this seems unlikely for higher distances, where
many of the young halo stars are observed. Both of these mechanisms
would produce a few or up to tens of young stars with correlated velocities
whereas formation of isolated stars is unlikely. In one case \citet{lyn+02}
have studied the environment of the halo young star PHL 346, but found
no evidence for similarly young stars in its vicinity. 

Rejuvenation scenarios of runaway stars have been discussed in the
literature, but in a different context, dealing with mass accretion
onto a star following a supernova explosion \citep{mar03}, and the
possible formation of a Thorne-Zytkow object \citep{leo+93}. These
scenarios dealt either with a specific observed runway \citep{mar03},
or with more rare cases than the scenario discussed here. They also
discussed interaction of a compact object with its companion rather
than the rejuvenation due to mass transfer in the post-main sequence
evolution stages of binaries. 

In the following we discuss the status of the candidate runaway and
hypervelocity stars in the Galactic halo and the implications of the
binary rejuvenation and the triple disruption scenarios for the young
stellar population in the GC. We suggest that all of the currently
observed young Halo stars could have been ejected from the Galactic
disk or the GC, when the binary rejuvenation scenario is taken into
account. We also expand our discussion on the origin of the hypervelocity
star HE 0437-5439, which was recently discussed in the literature.
We show that an ejection origin by an IMBH or through interactions
in a massive cluster in the LMC \citep{gua+07,gva+07} are highly
unlikely to be the origin of this star and suggest this HVS as a candidate
rejuvenated HVS ejected as a binary from the GC, which later on merged
to form a more massive HVS. We also show that its observed low metallicity
is a consistent with a Galactic origin and a GC origin can not be
ruled out.

\subsection{Rejuvenated runaway stars}

In table 1 we have listed young stars observed in the Galactic halo
that have estimated propagation times which could potentially be larger
than their evolutionary time, given the uncertainties. As we have
shown in section \ref{sec:rej}, the rejuvenation in binaries can
extend the travel time of the runaway stars as main sequence stars.
Following the rejuvenation, in both the full merger and the strong
mass transfer cases discusses above, we find that the newly rejuvenated
massive star, with mass $m_{rej}$, contains most of the mass of the
initial binary. The typical formation timescale is of the order of
the MS lifetime of the prime binary component, $t_{m_{1}}$. If such
binaries are ejected as runaway or hypervelocity stars they could
propagate for as long as $t_{m_{1}}$ before producing the newly formed
massive star. The MS lifetime of this rejuvenated star, $t_{mrej}$,
could then be much smaller then the propagation time, $t_{mrej}<t_{prop}\le t_{m_{1}}$,
thus producing an apparent discrepancy between the flight time and
the lifetime of this star. 

Such rejuvenation scenarios provide a maximal flight time for a given
star of $t_{prop}=t_{m_{1}}+t_{mrej}$. Since the mass of the prime
component in the binary progenitor is at least half the mass of the
rejuvenated star we find the maximal propagation time to be $t_{prop}^{max}=t_{\frac{1}{2}mrej}+t_{mrej}$,
where $t_{\frac{1}{2}mrej}$ is the MS lifetime of a star with half
the mass of the observed halo star. We find that all of the candidate
stars in table 1 have $t_{flight}<t_{prop}^{max}$, and could be rejuvenated
stars.

\subsection{Rejuvenated hypervelocity stars and the case for HE 0437-5439 and
US 708}

Currently \textasciitilde{}20 HVSs have been observed in the Galactic
halo \citep{bro+05,hir+05,ede+06,bro+07a,bro+07b}. The kinematics
and ages of most of these stars are consistent with their possible
origin from the GC. A discrepancy between the kinematics and the ages
of a few of the bound HVSs might exist \citep{bro+07a}, making them
possible candidate rejuvenated stars. However, more observations are
required to confirm that these stars are truly early B-type stars,
and not halo extreme horizontal branch stars. 

The hypervelocity star HE 0437-5439 was spectroscopically identified
to be a genuinely young B2 III-IV halo star with mass of $\sim9\pm0.8\, M_{\odot}$
\citep{ede+06,bon+08,prz+08}. Its apparently short lifetime and the
large distances from the GC and the Galactic disk make it too young
to have traveled from these regions during its lifetime, even with
it very high velocity ($723\, km\, s^{-1}$;\citealp{ede+06})%
\footnote{Recently, and after this study was done, another HVS was found suggesting
an origin outside the GC \citep{heb+08}. We do not discuss this interesting
star in the current paper and leave it to future study.%
}. It was therefore suggested to be either ejected from the large Magellanic
cloud (LMC) or rejuvenated in a binary ejected by an IMBH inspiral
to the GC \citep{ede+06}. In the following we discuss the possible
evidence for the origin of this star (metallicity), and its dynamical
history. We show that the suggested dynamical origins of this star
from the LMC require improbable dynamical scenarios, and as an alternative
we suggest it is a rejuvenated star ejected as a hypervelocity binary
in a triple disruption by the MBH in the GC. We also show that the
chemical abundances of HE 0437-5439 suggested as evidence for an LMC
origin of the HVS, do not rule out its possible Galactic center origin,
and could be consistent with such a scenario. In addition we shortly
discuss the possible rejuvenation origin of the low mass old HVS US
708.

\subsubsection{Ruling out some possible dynamical origins of HE 0437-5439 from the LMC}

Given the observed high velocity of HE 0437-5439, a dynamical scenario
for this star would most likely require an interaction with a MBH.
\citet{gua+07} have suggested that a tidal disruption of a binary
by an IMBH in a young stellar cluster in the LMC could produce hypervelocity
star such as HE 0437-5439 at a rate of $5\times10^{-8}\, yr^{-1}$.
We note that such IMBH have not yet been observed in the LMC (or elsewhere).   
Moreover, \citet{gua+07} have not taken into account a few important
considerations. (1) The travel time from the LMC to our galaxy for
such a HVS is about 20 Myrs, very close to the ages of the clusters
they suggested as possible hosts of an IMBH. For these clusters the
HVS should have been ejected immediately after the formation of the
IMBH (assuming the IMBH have formed quickly enough in the cluster
to begin with), probably during less than 1 Myrs, in order to achieve
its current position. Given the calculated ejection rates, only $N_{eject}\sim10^{6}\times5\times10^{-8}=0.05$
HVSs could have been ejected, on average, in the relevant time. (2)
The stellar mass function has not been taken into account by \citet{gua+07}.
The fraction of stars as massive as $8.5\, M_{\odot}$ or more (the
estimated mass of HE 0437-5439) is very small, only a fraction of
$f_{IMF}\sim0.01$ of the stellar population is in such massive stars
(and likely even smaller, as most of these more massive stars were likely to 
fuel the growth of the IMBH).
(3) The ejection of HVSs is isotropic, and therefore only a fraction
of them $f_{MW}<0.1$ would be directed to the Milky Way galaxy. Taking
together the average number of observable massive ($>8.5\, M_{\odot}$)
HVSs from the LMC (similar to HE 0437-5439) should be about $N_{eject}\times f_{imf}\times f_{MW}=5\times10^{-5}$,
making this possibility highly unlikely. 

Recently it was suggested that HVSs might be produced through binary-binary
dynamical interactions of massive binaries in a dense cluster \citep{gva+07}.
\citet{leo+91} have studied such encounters. He found that the lowest
mass star participating in the interaction could attain the highest
velocity. Such velocity would be comparable to the escape velocity
from the the most massive star participating in the encounter. If
HE 0437-5439 was ejected from the LMC, it would require an ejection
velocity of $>900\, km\, s^{-1}$ in order to acquire its current
position during its lifetime \citep{gua+07,prz+08}. For this to happen
HE 0437-5439 would need to encounter stars more massive than itself.
Even then the fraction of encounters where such velocity could be
attained by this star is $f_{high}\sim10^{-4}\,(4\times10^{-4},2\times10^{-3}$;
see \citealt{leo+91}) for encounters where the masses of the other
stars are larger than $15\, M_{\odot}(30\, M_{\odot},60\, M_{\odot}$;
respectively). Since binary-binary encounters usually lead to the
disruption of one of the binaries, one would require $\sim1/f_{high}$
such binaries to exist in order for one of them to potentially be
ejected at such high velocity. Such conditions, i.e. the existence
of hundreds (thousands) of $>30\, M_{\odot}(>15\, M_{\odot})$ stars
in a super dense cluster core are not known to exist in any young
cluster in the Galaxy or in the LMC. We conclude that the scenario
for ejection of HE 0437-5439 through a dynamical interaction with
massive stars in a cluster is highly unlikely.

\subsubsection{The metallicity of HE 0437-5439 does not rule out a Galactic center
origin}

Recently \citet{bon+08} and \citet{prz+08} have found the metallicity
of the HVS HE 0437-5439 to be low relative to solar metallicity. They
suggested this as a possible evidence for an LMC origin of this star
rather than a Galactic origin. However, an LMC origin would be difficult 
to explain dynamically, as discussed above. Moreover, in the following we 
show that the current metallicity measurements do not rule out a Galactic 
(and Galactic center) origin for HE 0437-5439. 

It is known that the observed abundances of some elements in Galactic
B stars are depressed relative to the established solar values (see
e.g. \citealp{mar04,mar06} and the appendix). Therefore, one should
compare the metallicity of B type stars such as HE 0437-5439 with
large surveys of similar B type stars. \citet{prz+08} made a comparison
with a single galactic B star and a single LMC B star that may not
representative of the large scatter in the abundances shown in larger
samples. \citet{bon+08} made a comparison with a large B stars survey
in the LMC, but for the comparison with the Galactic abundances they
took the solar abundances rather that Galactic B stars surveys. 
Fig. 2 show the spread in the measured chemical abundances for different 
samples of stars (LMC, milky way and the Galactic center stars) found in 
the literature (see caption of fig. 2).
For small samples (with 10 stars or less) the data for each 
of the stars is shown rather than the mean abundance,
since given the very small samples the mean may not be a good representative
of the underlying distribution of abundances. For the larger samples
the 1 $\sigma$ spread around the mean (i.e. where ~68 percents of the 
measured values are found) is shown \footnote{Notice that the specific 
Fe abundance  for each of the LMC stars is not given by Hunter et al. (2007), 
and we therefore quote the mean and uncertainty given by them in this case.}
We can use these data samples and compare them with the chemical abundances of 
HE 0437-5439. The results by \citet{prz+08} have systematically smaller error
bars, and also contain the abundances for more elements 
($C,\, N,\, O,\, Mg,\, Si$ and $Fe$) compared with the results obtained by \citet{bon+08} (which do not show the Fe abundance), and ere therefore used in 
the comparison. 
Nevertheless, given the large differences
that exist between the chemical abundances obtained by \citet{prz+08}
and those found by \citet{bon+08}, the latter results are also shown in fig. 2
for completeness. 

Comparison of the chemical abundances of HE 0437-5439 as found by 
\citet{prz+08} to those found in 
surveys of B type stars throughout the Galaxy (\citealp{daf+01}
show that its metallicity is highly consistent
with their metallicities%
\footnote{Even in such surveys large uncertainties exist; in the appendix we
show the chemical abundances of B stars found both in the Galaxy and
the LMC in several different surveys, enabling a more detailed comparison.
Fig. 2 shows only the results from the survey of \citealp{daf+01},
nevertheless, this survey includes B stars from different regions
in the Galaxy, and it is generally consistent with the other samples
(see appendix). One could also compare the values obtained for HE
0437-5439 to those of other young halo stars showing that its metallicities
are not unusual for such objects. %
}. Most (>50 percents) of the stars in the sample have more extreme elemental 
abundances than that of HE 0437-5439 
(for each of the elemental abundances observed; see fig. 2, bottom panel), 
showing that the elemental abundances 
of HE 0437-5439 are quite typical of the Milky Way chemical abundances.

Although a galactic origin is most consistent with the metallicities of
HE 0437-5439, its high velocity would require an interaction with
a MBH, currently known to exist only in the GC. Therefore, its metallicities
should be compared to the metallicities of similar unevolved B stars
in the GC. Unfortunately, metallicity measurements of stars in the
GC region exist only for a small number of stars, none of which are
similar to HE 0437-5439 (although some of these GC stars have similar
masses, they are at a very different evolutionary stage; \citealt{cun+07}).
Therefore, given the current data, drawing conclusions on the origin
of HE 0437-5439 based on metallicity comparisons with GC stars is 
premature. Nevertheless, we shortly discuss such metallicity comparison,
but caution that this should not be taken as evidence for the origin
of HE 0437-5439, but at most as a possible clue until further measurements
of the metallicity of GC stars are available. 

The chemical abundances of stars in the GC are known mostly for cool
evolved stars \citep{cun+07}. Data on two additional highly massive
($\sim150\, M_{\odot}$) LBV stars exists \citep{naj+08}, but given
their very different stellar type and evolution, comparing their abundances
with that of HE 0437-5439 is not justified, and we do not use their
data. The number of data points (stars) for each element are 10, 7,
6 and 5 for the elements $Fe,\, O,\, C$ and $N$, respectively. 
The $C,\, O$ and $Fe$ abundances of HE 0437-5439 are found 
to be consistent with GC values (see fig. 2%
\footnote{\citet{prz+08} have shown a somewhat similar plot, however the data
shown by \citet{prz+08} for the GC stars did not include IRS 8 \citep{geb+06}
and some of the elemental abundances for IRS 7 \citep{car+00}, that
show a larger scatter in the element abundances. They also used, with
no justification, the data from the GC LBV stars.}. 
), but the $N$ abundance is not (with {\bf all} 
5 stars in the GC sample have higher abundances than those
found for HE 0437-5439).  We note that better
agreement is observed for those elements for which more data exist.
The $N$ abundance of HE 0437-5439 is lower than those of the GC stars.
However, since CNO cycle mixing can convert these elements in the
GC stars (see e.g. \citealt{car+00,cun+07}) and in HE 0437-5439,
one should be careful and also check the sum of these elements and
not only compare each of these elements by itself. We find that the
sum of these elements is consistent with that of the GC sample stars,
i.e. not even the $N$ abundance of HE 0437-5439 could be interpreted
as an evidence against a GC origin. Moreover, given the low statistics
of the GC sample, the lower N abundance is not statistically significant,
even by itself. Given the wide range of abundances found in different
Galactic B-type stars surveys (see appendix), it is possible that
this specific HVS B star may have lower abundances of these specific
element. 

The middle panel of fig. 2 shows the comparison of the elements abundances
of HE 0437-5439 to those found in the LMC. The metallicities of B
stars in the LMC were found by several groups (the largest samples
by \citealp{kor+02,hun+07}), where we show the values obtained by
\citeauthor{hun+07}, for which a large sample exists (30), whereas
the sample by \citeauthor{kor+02} contain only 4 stars).
The comparison shows much poorer agreement with LMC abundances than with 
the Milky way abundances.
The N, Fe and the Si abundances of HE 0437-5439 are found 
to be consistent with LMC values, but the $C,\, O$ and $Mg$ abundances are not
({\bf all} 30 stars in the LMC samples have higher abundances than those
found for HE 0437-5439 for these elements).
We do note, however, that the metallicities
of HE 0437-5439 obtained by \citet{bon+08} 
are consistent with those of the LMC for all elements. In addition
the results of \citet{prz+08} are better consistent with LMC values 
found by \citeauthor{kor+02} which are systematically higher). 

These comparisons show that the metallicities obtained
by \citet{prz+08} are more consistent with the GC abundances than
the LMC abundances shown here (and best consistent with a Galactic
origin). However, we caution again that the small statistics and the
different type of stars included in the GC sample of stars, the large
differences inferred for the metallicities of HE 0437-5439 by different
authors \citep{bon+08,prz+08}; and the different analysis methods
used by different groups, suggest that it is still to early to draw
conclusions from such metallicity comparisons. It is clear from the
above discussion, however, that none of the suggested origins 
for HE 0437-5439, including the Galactic disk, the LMC and the GC 
can be ruled out based upon current metallicity data.  

\begin{figure}
\includegraphics[scale=0.3]{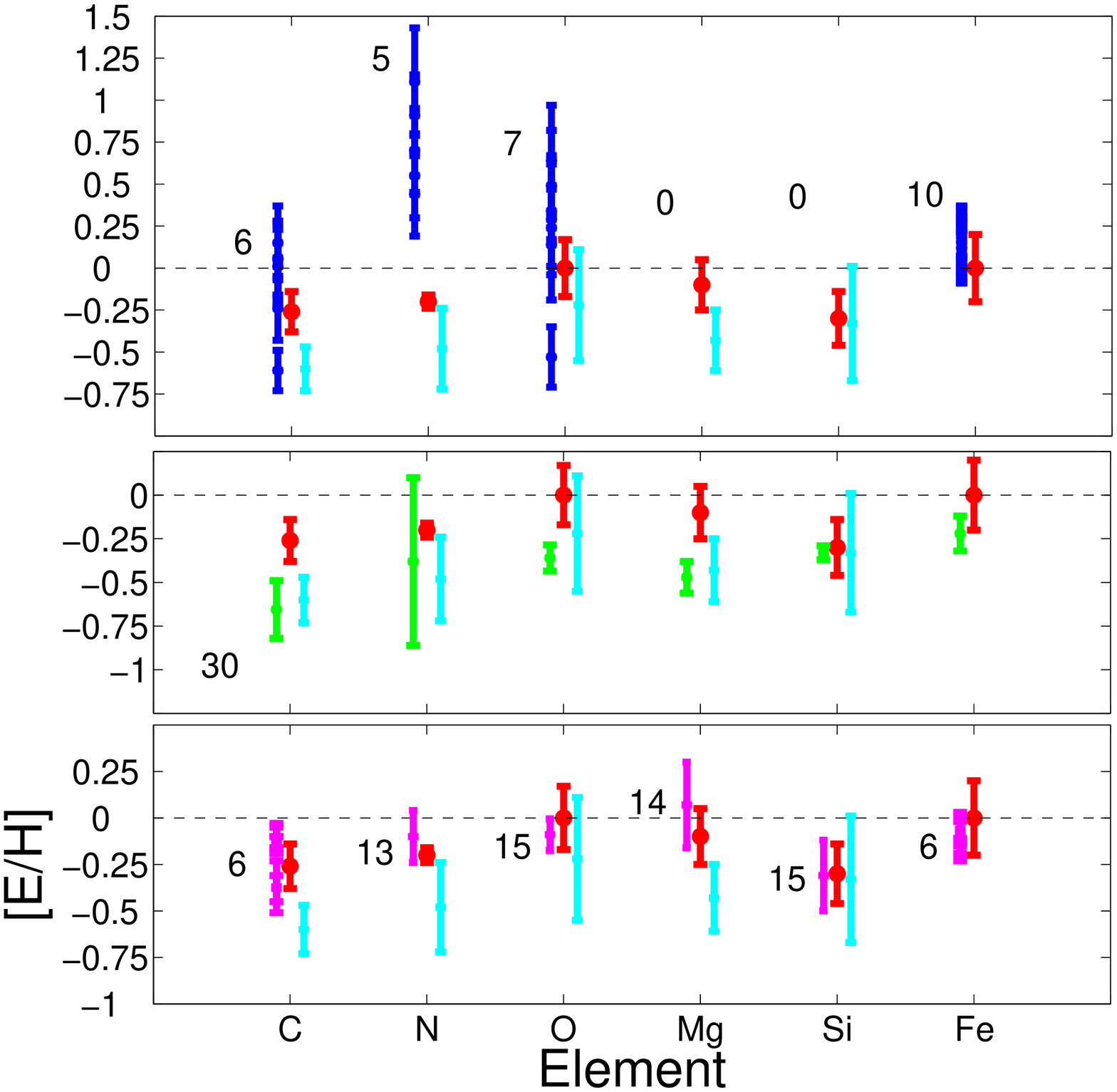}

\caption{\label{fig:metallicities}Comparison of the metal abundances of the
hypervelocity star HE 0437-5439 (\citealp{prz+08} middle red bars
and \citealp{bon+08} right cyan bars) to those of the Galactic center
(top panel, left blue bars), LMC B stars (middle panel, left green
bars) and Galactic B stars (bottom panel, left magenta bars). Galactic
B stars metallicities are taken from \citet{daf+01}, LMC B stars
metallicities are taken from \citet{hun+07}. Galactic center metallicities
are taken from \citealp{car+00,geb+06,cun+07,naj+08}. All values
are given relative to solar abundances \citep[dashed lines;][]{gre+07}
. Note that the GC metallicity values are shown for each of the stars
in those smaples containing ten or less stars (see text), whereas
for larger sample sizes the median metallicity values and the 1 $\sigma$ spread around the median are shown. 
The number of stars from which the range of abundance values were obtained is denoted beside
the bars (the LMC samples includes 30 values for all the elements beside $Fe$, see text.}

\end{figure}

\subsubsection{The possibility of HE 0437-5439 as a rejuvenated star from the Galactic
center}

As discussed in the previous sections, young halo stars such as HE
0437-5439 could have evolved from two lower mass members of ejected
hypervelocity binaries. The binary components either merge completely,
or may evolve through a strong mass transfer from the more massive
binary component to its initially lower mass companion, which becomes
a rejuvenated massive star. The later possibility would give a strong
observable signature in the form of a high mass ratio binary with
a period of tens of days. The observations of \citet{prz+08} and
\citet{bon+08} rule out the possibility of such a binary, and therefore
if this HVS is a rejuvenated star its binary progenitor fully merged.
As can be seen in fig. \ref{fig:Evolutionary-time-scale}, a binary
progenitor with $4.8$ and $4.3\, M_{\odot}$ components, for example,
could have propagated for more than $60-80$ Myrs and then merge to
form a $9\, M_{\odot}$ star , consistent with the required travel
time from the GC \citep{ede+06}. We note that binaries with other
high mass ratio components (not shown here), but with a total binary
mass of $\sim9\, M_{\odot}$ could also produce such a merged star
with the appropriate evolutionary time scales. Given the tendency
of massive binaries to have high mass ratios, it is likely that most
ejected binaries should have mass ratios in the relevant parameter
space as to produce a merged star similar to HE 0437-5439. All the
hypervelocity binaries have initial semi-major axis smaller than $\sim0.2\, AU$
($4-5\, R_{\odot}$), since they were originally part of the inner
binaries in close stable triples. In addition, the KCTF mechanism
in triples (see section 4) can evolve such inner binaries into even
closer contact configuration at a very short timescale (Myr). It is
therefore expected that most hypervelocity binaries would be at very
tight orbits or a few $R_{\odot}$ separations, such as those required
to eventually form a merged star like HE 0437-5439 (see fig. 1). In
other words the rejuvenation scenario is a likely scenario for the
formation and ejection of HE 0437-5439. The estimate in section 3.2.1
suggests $f_{prog}=3-5\%$ of the HVSs could be ejected as close hypervelocity
binaries or leave a binary star close to the MBH. The fraction of
rejuvenated HVSs is therefore $f_{rej}=f_{prog}\cdot f_{inner}\cdot f_{merged}\cdot f_{lifetime}$
where $f_{inner}$ is the fraction of disrupted triples that eject
the inner binary, $f_{merger}$ is the probability the binary merges,
$f_{lifetime}$ is the fractional lifetime of the merger remnant to
the original stars lifetime. Simulations of binary disruptions by
a MBH \citep{mil+05,bro+06c} suggest that there is no preference 
for the primary or the secondary in the binary to be ejected, i.e. they
have equal ejection probability, I therefore take $f_{inner}=0.5$. Estimating
the probability of binary merger is very uncertain. Taking the observed
triples sample of \citet{fek81}, I find that the inner binaries in
the HVSs binaries possible progenitors have periods of a few days,
which together with the binary stellar evolution simulations (see
fig. \ref{fig:Evolutionary-time-scale}) suggest that $> 0.75$ of
them would merge due to stellar evolution. I therefore take $f_{merger}=0.75$.
The question of $f_{lifetime}$ is more complicated. The merger remnant
has a lifetime of $t_{mrej}$ as discussed in section 5.1, and the
maximal lifetime of its progenitor is at most $t_{\frac{1}{2}mrej}$.
A rejuvenated star therefore has a fractional life time to the original
stars of $t_{mrej}/t_{\frac{1}{2}mrej}$. Taking the stellar evolution
main sequence lifetimes from the Geneva stellar evolution tracks \citep{sch+92a},
I find this to be approximately $0.2$. However, on average, the HVSs
are ejected after half their lifetime. In addition, the ejected stars
are observed only after their propagation for $t_{prop}=few\times10^{7}-10^{8}$
yrs (the flight time to distances of $10-100$ kpc from the GC where
they are currently observed), the fraction of the merger remnant lifetime
to the time spent by the progenitor at observable regions is $t_{mrej}/(0.5\cdot t_{\frac{1}{2}mrej}+t_{prop})>0.4$.
A reasonable estimate is therefore $f_{lifetime}=0.2-0.6$. Taken
together, we find $f_{rej}\simeq0.003-0.012$%
\footnote{One could also discuss the probability of getting a metallicity distribution
such as observed for HE 0437-5439 given the metallicities observed
in the GC. As discussed above, its metallicity is consistent with
the GC metallicities of similar massive stars. However, current data
can not be reliably used regarding this point, since these stars are
at different evolutionary stages and are not comparable to HE 0437-5439.%
}. Given the current estimate of $N_{HVS}\sim100-650$ $3-4\, M_{\odot}$
B type HVSs in the galactic halo \citet{bro+07b,per+08b}, we may
expect to find $N_{HVS}\cdot f_{rej}=0.25-8$ rejuvenated (more) massive
HVSs in a full sky survey (compare with the value of $5\times10^{-5}$
massive stars expected to originate from the LMC in the IMBH scenario,
if such IMBH indeed exist there in a young massive cluster). 

We summarize this section by concluding that current suggestions for
the hypervelocity ejection of HE 0437-5439 from the LMC (either by
an IMBH or through interactions with other massive stars) are highly
unlikely, and suggest that HE 0437-5439 was ejected from our GC (where
we stress that this scenario is not ruled out by current metallicity
comparisons) in a hypervelocity binary following a triple disruption
by the MBH.

\subsubsection{The possibility of US 708 as a rejuvenated star from the Galactic
center}

We note that the hypervelocity star US 708, which is not a young star
but an evolved sdO star, was suggested to evolve from a merger of
a close binary \citep{hir+05}. In that respect we find that when
the evolution of close binaries such as studied here is followed to
later times (not shown here) such white dwarfs are indeed formed in
some cases. It is therefore possible that US 708 was also ejected
as a binary star from a triple disruption that later on evolved and
merged to form an sdO star, in a similar way to the rejuvenated young
halo star studied here (as suggested by W. Brown, private communication).
However, the unique evolution and the a merger of two white dwarfs
is only poorly followed in the evolutionary code used here, and further
studies should be made to check this possibility.

\subsection{Binaries in the S-stars cluster in the Galactic center}

In recent years high resolution observations have revealed the existence
of many young OB stars in the GC near the MBH. Most of the young stars
are observed in the central 0.5 pc around the MBH. The young stars
population in the inner 0.04 pc (the 'S-stars') contain only young
B-stars, in apparently isotropic distribution around the MBH \citep{eis+05,ghe+05}.
The young stars outside this region contain many O-stars in a disk
like structure and were probably formed from the fragmentation of
a gaseous disk \citep{lev+03,pau+06}. However, the origin of the
S-stars is difficult to explain by this process. It was suggested
that the S-stars with their very different properties migrated from
the stellar disk through a planetary migration like process \citep{lev07}.
This interesting possibility has not yet been studied quantitatively.
Another possibility is that these stars have a different origin, possibly
from the disruption of young binaries and the following capture of
one of their components \citep{gou+03}. It was recently shown that
such a scenario could be consistent with the current knowledge regarding
the number of the observed S-stars and their orbital properties \citep{per+07}.
In case of a triple disruption, a binary could be captured at a close
orbit near the MBH. We therefore suggest that some of the observed
S-stars may be binaries. Such binaries may not survive for long in
this environment due to interactions with other stars \citep[See ][ for a detailed discussion on binaries survival in the GC]{per09}.
Such interactions could also change the orbits of the disrupted binary
components near the MBH. Still, the closest binaries could survive
for longer times \citep{per09} and may be observed. Interestingly,
observations of wider S-stars binaries would suggest more recent capture,
i.e. the binary period may serve to estimate the time a binary spent
in the close by regions of the MBH. In addition such S-star binaries
could rejuvenate to form more massive S-stars, possibly explaining
the occurrence of the most massive S-stars such as S-2.

\section{Summary and conclusions}

In this paper we have studied a possible explanation for the existence
of young stars far in the Galactic halo. Such stars are usually thought
to have formed elsewhere, either in the Galactic disk or the Galactic
center, and later on ejected at high velocities into their current
position. However, some of these stars have apparent lifetimes shorter
the required flight time from the Galactic disk/center. Here we suggested
that such stars have evolved in close runaway or hypervelocity binaries.
We found that stellar evolution of such binaries can drive them into
mass transfer configurations and even mergers. Such evolution could
then rejuvenate them (similar to the cases of lower mass blue stragglers)
and extend their lifetimes after ejection. The extended lifetimes
of such stars could then be reconciled with their flight times to
the Galactic halo, and the travel times could be extended up to 3-4
times relative to their apparent lifetimes. 

Three typical scenarios were found for the binaries evolution. (1)
In case of a full merger of the binary progenitor components, a single
halo star would be observed. Unless the binary merger lead to peculiar
characteristics of the merged star, such as possibly high rotational
velocity , or chemical peculiarities, observation of such stars could
not directly trace their binary origin. Nevertheless, we predict that
in such cases the calculated propagation time of the star from its
birthplace would be limited to approximately the main sequence lifetime
of its binary progenitor components (at most the main sequence lifetime
of a star with half the mass of the observed halo star). (2) In the
case of a strong mass transfer, one of the binary progenitor component
have accreted most of the mass of its companion. In such a case the
evolved binary could be observable as a high mass ratio binary, with
typical periods of $10-200$days. (3) In the case of a weak mass transfer,
only a small fraction of the mass of the binary progenitor component
is accreted by its companion, leading to negligible or mild rejuvenation.
This third evolutionary route would not produce {}``too young''
halo stars, albeit they could produce marginal cases. However, such
halo young binaries (that have already been observed) can confirm
the scenario of runaway binaries that, for different orbital parameters
of the binary, could have produced rejuvenated high velocity stars. 

We studied the possibilities of binary runaway and hypervelocity stars
and showed that such binaries could have been ejected in triple disruptions
and other dynamical interactions with stars or with massive black
holes. Consequently, currently observed {}``too young'' star in
the halo could have been ejected from the Galactic disk or the Galactic
center and be observable in their current position if they were ejected
as binaries (whereas other suggestions such as ejection from the LMC
are shown to be highly unlikely). The calculated propagation times
of these stars are indeed consistent with the evolutionary lifetimes
in rejuvenated binaries. We suggest to look for binary companions
that may exist for some of these stars, thus directly confirming the
binary rejuvenation scenario (HD 188618 in table 1 has already been
suggested as binary candidate; \citealt{mar06}). We also specifically
discuss the hypervelocity star HE 0437-5439 in that respect, and show
that it could be a rejuvenated HVS from the GC (where we also discuss
its recently observed metallicity, showing that a Galactic center
origin can not be currently ruled out, contrary to recent suggestions)
.

Finally, we also suggest that triple disruptions by the massive black
hole in the Galactic center could capture binaries in close orbits
near the MBH, some of which may later evolve to become more massive
rejuvenated stars. Future observations might be able to study these
binaries.

\acknowledgements{I would like to thank Warren Brown, Uli Heber, Clovis Hopman, Tal
Alexander, Mercedes Lopez-Morales and the anonymous referee for helpful
comments and suggestions on this manuscript. I would also like to
thank Daniel Fabrycky for using his code for producing stable triples
for my calculations of the hypervelocity binaries fraction, and Ben
Davies for helpful correspondence.}

\appendix

\section{A. Chemical abundances of young B stars and halo stars}

Following section 5.2.2 we supplement in this appendix a table in
which the detailed chemical abundances of B stars found in Galactic
and LMC surveys are shown, in order to give better and broader perspective
on their scatter and uncertainties. These show the typically lower
measured metallicities characterizing B stars, in comparison with
the solar abundances. Also shown are the chemical abundances found
for HE 0437-5439 by \citet{bon+08} and \citet{prz+08}. In addition
we detail the metallicities of other candidate rejuvenated halo stars
from table 1 (for which measurements are available), showing that
the metallicities of HE 0437-5439 are not unusual. The quoted
mean abundances in the surveys have typical $0.1-0.2$ dex uncertainties
in the mesurements. The specific uncertainties for single stars is given 
specifically for each star.

\begin{table}
\caption{\label{t:chem2}Chemical abundances of B stars}

\begin{centering}
\begin{tabular}{lcccllll}
\hline 
Name  & $C$ & $N$ & $O$ & $Mg$ & $Si$ & $Fe$ & \tabularnewline
\hline 
LMC B1-2 stars  & $7.73$ & $6.88$ & $8.33$ & $7.06$ & $7.19$ & $7.23$ & \citet{hun+07}\tabularnewline
LMC B V stars & $8.06$ & $7.01$ & $8.37$ & $7.37$ & $7.1$ & $7.33$ & \citet{kor+02}\tabularnewline
Solar  & $8.39$ & $7.78$ & $8.66$ & $7.53$ & $7.51$ & $7.45$ & \citet{gre+07}\tabularnewline
MW B1-2 III/IV stars & $7.82$ & $7.38$ & $8.68$ & $7.09$ & $7.14$ & -- & \citet{vra+00}\tabularnewline
MW B IV/V stars & $8.22$ & $7.78$ & $8.52$ & $7.38$ & $6.81$ & -- & \citealp{kil92}\tabularnewline
{}`` & $8.20$ & $7.81$ & $8.68$ & -- & $7.58$ & $7.72$ & \citealp{gie+92}\tabularnewline
{}`` & $8.24$ & $7.69$ & $8.6$ & $7.65$ & $7.25$ & $7.35$ & \citealp{daf+01}\tabularnewline
MW B stars & $8.21$ & $7.98$ & $8.55$ & $7.56$ & $7.36$ & $7.4$ & \citealp{mar04}\tabularnewline
 & $8.29$ & $7.9$ & -- & -- & -- & -- & \citealp{nie+06,nie+07}\tabularnewline
\hline 
HE 0437-5439 (B2 III-IV star) & $8.13\pm0.12$ & $7.58\pm0.04$ & $8.66\pm0.17$ & $7.43\pm0.15$ & $7.21\pm0.16$ & $7.45\pm0.2$ & \citet{prz+08}\tabularnewline
HE 0437-5439 & $7.79\pm0.13$ & $7.3\pm0.24$ & $8.44\pm0.33$ & $7.1\pm0.18$ & $7.18\pm0.34$ & -- & \citet{bon+08}\tabularnewline
\hline
BD +38 2182 & -- & -- & -- & -- & $7.02\pm0.04$ & -- & \citealt{mar04}\tabularnewline
BD +36 2268 & $7.14\pm0.19$ & $7.89\pm0.2$ & $8.27\pm0.37$ & -- & $6.6\pm0.2$ & -- & {}``\tabularnewline
HD 140543 & -- & $8.18$ & $8.76\pm0.46$ & -- & $8.74\pm0.43$ & -- & {}``\tabularnewline
HD 188618 & -- & $7.93\pm0.00$ & $8.51\pm0.23$ & -- & $7.71\pm0.35$ & -- & {}``\tabularnewline
HD 206144 & -- & $7.86$ & $8.58\pm0.33$ & -- & $7.56\pm0.15$ & $6.99$ & {}``\tabularnewline
PHL 159 & $8.17\pm0.26$ & $7.85\pm0.18$ & $8.72\pm0.16$ & $7.28\pm0.2$ & $7.34\pm0.2$ & $7.33\pm0.09$ & {}``\tabularnewline
PHL 346 & $8.14\pm0.32$ & $8.04\pm0.19$ & $8.54\pm0.28$ & $7.14$ & $7.47\pm0.08$ & $7.38\pm0.16$ & {}``\tabularnewline
PG 1209+263 & -- & -- & -- & $5.03$ & $8.31\pm0.04$ & $7.22\pm0.56$ & \citealt{lyn+04}\tabularnewline
PG 2219+094 & $7.82$ & -- & -- & $7.16$ & -- & -- & \citealt{rol+99}\tabularnewline
PG 2229+099 & $8.15\pm0.1$ & $7.94\pm0.31$ & $8.99\pm0.06$ & $7.1\pm0.62$ & $7.36$ & $5.63$ & {}``\tabularnewline
\hline
\end{tabular}
\par\end{centering}
\end{table}

\newpage{}


\begin{thebibliography}{106}
\expandafter\ifx\csname natexlab\endcsname\relax\def\natexlab#1{#1}\fi

\bibitem[{{Abt}(1983)}]{abt83}
{Abt}, H.~A. 1983, \araa, 21, 343

\bibitem[{{Abt} {et~al.}(1990){Abt}, {Gomez}, \& {Levy}}]{abt+90}
{Abt}, H.~A., {Gomez}, A.~E., \& {Levy}, S.~G. 1990, \apjs, 74, 551

\bibitem[{{Baumgardt} {et~al.}(2006){Baumgardt}, {Gualandris}, \& {Portegies
  Zwart}}]{bau+06}
{Baumgardt}, H., {Gualandris}, A., \& {Portegies Zwart}, S. 2006, \mnras, 372,
  174

\bibitem[{{Blaauw}(1961)}]{bla61}
{Blaauw}, A. 1961, \bain, 15, 265

\bibitem[{{Bonanos} {et~al.}(2008)}]{bon+08}
{Bonanos}, A.~Z. {et~al.} 2008, \apjl, 675, L77

\bibitem[{{Bromley} {et~al.}(2006)}]{bro+06c}
{Bromley}, B.~C. {et~al.} 2006, \apj, 653, 1194

\bibitem[{{Brown} {et~al.}(2005)}]{bro+05}
{Brown}, W.~R. {et~al.} 2005, \apjl, 622, L33

\bibitem[{{Brown} {et~al.}(2007{\natexlab{a}})}]{bro+07a}
---. 2007{\natexlab{a}}, \apj, 660, 311

\bibitem[{{Brown} {et~al.}(2007{\natexlab{b}})}]{bro+07b}
---. 2007{\natexlab{b}}, ArXiv:0709.1471

\bibitem[{{Carr} {et~al.}(2000){Carr}, {Sellgren}, \& {Balachandran}}]{car+00}
{Carr}, J.~S., {Sellgren}, K., \& {Balachandran}, S.~C. 2000, \apj, 530, 307

\bibitem[Chen \& Han(2004)]{che+04} 
Chen, X., \& Han, Z.\ 2004, \mnras, 355, 1182 

\bibitem[{{Christodoulou} {et~al.}(1997){Christodoulou}, {Tohline}, \&
  {Keenan}}]{chr+97}
{Christodoulou}, D.~M., {Tohline}, J.~E., \& {Keenan}, F.~P. 1997, \apj, 486,
  810

\bibitem[{{Cunha} {et~al.}(2007){Cunha}, {Sellgren}, {Smith}, {Ramirez},
  {Blum}, \& {Terndrup}}]{cun+07}
{Cunha}, K., {Sellgren}, K., {Smith}, V.~V., {Ramirez}, S.~V., {Blum}, R.~D.,
  \& {Terndrup}, D.~M. 2007, \apj, 669, 1011

\bibitem[{{Daflon} {et~al.}(2001)}]{daf+01}
{Daflon}, S. {et~al.} 2001, \apj, 552, 309

\bibitem[{{D'Angelo} {et~al.}(2006){D'Angelo}, {van Kerkwijk}, \&
  {Rucinski}}]{dan+06}
{D'Angelo}, C., {van Kerkwijk}, M.~H., \& {Rucinski}, S.~M. 2006, \aj, 132, 650

\bibitem[{{Dray} \& {Tout}(2007)}]{dra+07}
{Dray}, L.~M. \& {Tout}, C.~A. 2007, \mnras, 376, 61

\bibitem[{{Duquennoy} \& {Mayor}(1991)}]{duq+91}
{Duquennoy}, A. \& {Mayor}, M. 1991, \aap, 248, 485

\bibitem[{{Dyson} \& {Hartquist}(1983)}]{dys+83}
{Dyson}, J.~E. \& {Hartquist}, T.~W. 1983, \mnras, 203, 1233

\bibitem[{{Edelmann} {et~al.}(2005)}]{ede+06}
{Edelmann}, H. {et~al.} 2005, \apjl, 634, L181

\bibitem[{{Eggleton} {et~al.}(1990){Eggleton}, {Fitchett}, \& {Tout}}]{egg+90}
{Eggleton}, P.~P., {Fitchett}, M.~J., \& {Tout}, C.~A. 1990, \apj, 354, 387

\bibitem[{{Eggleton} \& {Kiseleva-Eggleton}(2001)}]{egg+01}
{Eggleton}, P.~P. \& {Kiseleva-Eggleton}, L. 2001, \apj, 562, 1012

\bibitem[{{Eggleton} {et~al.}(1989){Eggleton}, {Tout}, \& {Fitchett}}]{egg+89}
{Eggleton}, P.~P., {Tout}, C.~A., \& {Fitchett}, M.~J. 1989, \apj, 347, 998

\bibitem[{{Eisenhauer} {et~al.}(2005)}]{eis+05}
{Eisenhauer}, F. {et~al.} 2005, \apj, 628, 246

\bibitem[{{Evans} {et~al.}(2005)}]{eva+05}
{Evans}, N.~R. {et~al.} 2005, \aj, 130, 789

\bibitem[{{Fabrycky} \& {Tremaine}(2007)}]{fab+07}
{Fabrycky}, D. \& {Tremaine}, S. 2007, \apj, 669, 1298

\bibitem[{{Fekel}(1981)}]{fek81}
{Fekel}, Jr., F.~C. 1981, \apj, 246, 879

\bibitem[Ferraro et al.(2006)]{fer+06} 
Ferraro, F.~R., et al.2006, \apjl, 647, L53

\bibitem[{{Garmany} {et~al.}(1980){Garmany}, {Conti}, \& {Massey}}]{gar+80}
{Garmany}, C.~D., {Conti}, P.~S., \& {Massey}, P. 1980, \apj, 242, 1063

\bibitem[{{Geballe} {et~al.}(2006)}]{geb+06}
{Geballe}, T.~R. {et~al.} 2006, \apj, 652, 370

\bibitem[{{Ghez} {et~al.}(2005)}]{ghe+05}
{Ghez}, A.~M. {et~al.} 2005, \apj, 620, 744

\bibitem[{{Gies}(1987)}]{gie87}
{Gies}, D.~R. 1987, \apjs, 64, 545

\bibitem[{{Gies} \& {Bolton}(1986)}]{gie+86}
{Gies}, D.~R. \& {Bolton}, C.~T. 1986, \apjs, 61, 419

\bibitem[{{Gies} \& {Lambert}(1992)}]{gie+92}
{Gies}, D.~R. \& {Lambert}, D.~L. 1992, \apj, 387, 673

\bibitem[{{Ginsburg} \& {Loeb}(2006)}]{gin+06}
{Ginsburg}, I. \& {Loeb}, A. 2006, \mnras, 368, 221

\bibitem[{{Gould} \& {Quillen}(2003)}]{gou+03}
{Gould}, A. \& {Quillen}, A.~C. 2003, \apj, 592, 935

\bibitem[{{Greenstein} \& {Sargent}(1974)}]{gre+74}
{Greenstein}, J.~L. \& {Sargent}, A.~I. 1974, \apjs, 28, 157

\bibitem[{{Grevesse} {et~al.}(2007){Grevesse}, {Asplund}, \& {Sauval}}]{gre+07}
{Grevesse}, N., {Asplund}, M., \& {Sauval}, A.~J. 2007, Space Science Reviews,
  130, 105

\bibitem[{{Gualandris} \& {Portegies Zwart}(2007)}]{gua+07}
{Gualandris}, A. \& {Portegies Zwart}, S. 2007, \mnras, 376, L29

\bibitem[{{Gualandris} {et~al.}(2004){Gualandris}, {Portegies Zwart}, \&
  {Eggleton}}]{gua+04}
{Gualandris}, A., {Portegies Zwart}, S., \& {Eggleton}, P.~P. 2004, \mnras,
  350, 615

\bibitem[{{Gvaramadze} {et~al.}(2007){Gvaramadze}, {Gualandris}, \& {Portegies
  Zwart}}]{gva+07}
{Gvaramadze}, V.~V., {Gualandris}, A., \& {Portegies Zwart}, S. 2007, ArXiv:
  astro-ph/0702735

\bibitem[{{Hansen} \& {Milosavljevi{\' c}}(2003)}]{han+03a}
{Hansen}, B.~M.~S. \& {Milosavljevi{\' c}}, M. 2003, \apjl, 593, L77

\bibitem[{{Heber} {et~al.}(2008)}]{heb+08}
{Heber}, U. {et~al.} 2008, \aap, 483, L21

\bibitem[{{Hills}(1988)}]{hil88}
{Hills}, J.~G. 1988, \nat, 331, 687

\bibitem[{{Hills}(1991)}]{hil91}
---. 1991, \aj, 102, 704

\bibitem[{{Hirsch} {et~al.}(2005)}]{hir+05}
{Hirsch}, H.~A. {et~al.} 2005, \aap, 444, L61

\bibitem[{{Hoogerwerf} {et~al.}(2000){Hoogerwerf}, {de Bruijne}, \& {de
  Zeeuw}}]{hoo+00}
{Hoogerwerf}, R., {de Bruijne}, J.~H.~J., \& {de Zeeuw}, P.~T. 2000, \apjl,
  544, L133

\bibitem[{{Hoogerwerf} {et~al.}(2001){Hoogerwerf}, {de Bruijne}, \& {de
  Zeeuw}}]{hoo+01}
---. 2001, \aap, 365, 49

\bibitem[{{Hunter} {et~al.}(2007)}]{hun+07}
{Hunter}, I. {et~al.} 2007, \aap, 466, 277

\bibitem[{{Keenan}(1992)}]{kee+92}
{Keenan}, F.~P. 1992, \qjras, 33, 325

\bibitem[{{Kilian}(1992)}]{kil92}
{Kilian}, J. 1992, \aap, 262, 171

\bibitem[{{Kiseleva} {et~al.}(1998){Kiseleva}, {Eggleton}, \&
  {Mikkola}}]{kis+98}
{Kiseleva}, L.~G., {Eggleton}, P.~P., \& {Mikkola}, S. 1998, \mnras, 300, 292

\bibitem[{{Kobulnicky} \& {Fryer}(2007)}]{kob+07}
{Kobulnicky}, H.~A. \& {Fryer}, C.~L. 2007, \apj, 670, 747

\bibitem[{{Korn} {et~al.}(2002)}]{kor+02}
{Korn}, A.~J. {et~al.} 2002, \aap, 385, 143

\bibitem[{{Kouwenhoven} {et~al.}(2007)}]{kou+07}
{Kouwenhoven}, M.~B.~N. {et~al.} 2007, ArXiv: 0707.2746

\bibitem[{{Kozai}(1962)}]{koz62}
{Kozai}, Y. 1962, \aj, 67, 591

\bibitem[{{Leonard}(1991)}]{leo+91}
{Leonard}, P.~J.~T. 1991, \aj, 101, 562

\bibitem[{{Leonard}(1995)}]{leo95}
---. 1995, \mnras, 277, 1080

\bibitem[{{Leonard} \& {Duncan}(1988)}]{leo+88}
{Leonard}, P.~J.~T. \& {Duncan}, M.~J. 1988, \aj, 96, 222

\bibitem[{{Leonard} \& {Duncan}(1990)}]{leo+90}
---. 1990, \aj, 99, 608

\bibitem[{{Leonard} {et~al.}(1993){Leonard}, {Hills}, \& {Dewey}}]{leo+93}
{Leonard}, P.~J.~T., {Hills}, J.~G., \& {Dewey}, R.~J. 1993, in Astronomical
  Society of the Pacific Conference Series, Vol.~45, Luminous High-Latitude
  Stars, ed. D.~D. {Sasselov}, 386--+

\bibitem[{{Levin}(2006)}]{lev05}
{Levin}, Y. 2006, \apj, 653, 1203

\bibitem[{{Levin}(2007)}]{lev07}
---. 2007, \mnras, 374, 515

\bibitem[{{Levin} \& {Beloborodov}(2003)}]{lev+03}
{Levin}, Y. \& {Beloborodov}, A.~M. 2003, \apjl, 590, L33

\bibitem[{{Lockman} {et~al.}(2007){Lockman}, {Blundell}, \& {Goss}}]{loc+07b}
{Lockman}, F.~J., {Blundell}, K.~M., \& {Goss}, W.~M. 2007, \mnras, 381, 881

\bibitem[{{L{\"o}ckmann} \& {Baumgardt}(2007)}]{loc+07a}
{L{\"o}ckmann}, U. \& {Baumgardt}, H. 2007, ArXiv:0711.1326

\bibitem[{{Lu} {et~al.}(2007){Lu}, {Yu}, \& {Lin}}]{luy+07}
{Lu}, Y., {Yu}, Q., \& {Lin}, D.~N.~C. 2007, \apjl, 666, L89

\bibitem[{{Lynn} {et~al.}(2002)}]{lyn+02}
{Lynn}, B.~B. {et~al.} 2002, \mnras, 336, 1287

\bibitem[{{Lynn} {et~al.}(2004)}]{lyn+04}
---. 2004, \mnras, 349, 821

\bibitem[{{Mardling} \& {Aarseth}(2001)}]{mar+01}
{Mardling}, R.~A. \& {Aarseth}, S.~J. 2001, \mnras, 321, 398

\bibitem[{{Martin}(2003)}]{mar03}
{Martin}, J.~C. 2003, \pasp, 115, 49

\bibitem[{{Martin}(2004)}]{mar04}
---. 2004, \aj, 128, 2474

\bibitem[{{Martin}(2006)}]{mar06}
---. 2006, \aj, 131, 3047

\bibitem[{{Martos} {et~al.}(1999){Martos}, {Allen}, {Franco}, \&
  {Kurtz}}]{mar+99}
{Martos}, M., {Allen}, C., {Franco}, J., \& {Kurtz}, S. 1999, \apjl, 526, L89

\bibitem[{{Mason} {et~al.}(1998)}]{mas+98}
{Mason}, B.~D. {et~al.} 1998, \aj, 115, 821

\bibitem[{{McSwain} {et~al.}(2007{\natexlab{a}})}]{mcs+07c}
{McSwain}, M.~V. {et~al.} 2007{\natexlab{a}}, \apj, 655, 473

\bibitem[{{McSwain} {et~al.}(2007{\natexlab{b}})}]{mcs+07a}
---. 2007{\natexlab{b}}, \apj, 660, 740

\bibitem[{{Mikkola}(1983)}]{mik83}
{Mikkola}, S. 1983, \mnras, 205, 733

\bibitem[{{Miller} {et~al.}(2005){Miller}, {Freitag}, {Hamilton}, \&
  {Lauburg}}]{mil+05}
{Miller}, M.~C., {Freitag}, M., {Hamilton}, D.~P., \& {Lauburg}, V.~M. 2005,
  \apjl, 631, L117

\bibitem[{{Miralda-Escud{\' e}} \& {Gould}(2000)}]{mir+00}
{Miralda-Escud{\' e}}, J. \& {Gould}, A. 2000, \apj, 545, 847

\bibitem[{{Morrell} \& {Levato}(1991)}]{mor+91}
{Morrell}, N. \& {Levato}, H. 1991, \apjs, 75, 965

\bibitem[{{Najarro} {et~al.}(2008)}]{naj+08}
{Najarro}, F. {et~al.} 2008, \apj accepted

\bibitem[Nelson \& Eggleton(2001)]{nel+01} 
Nelson, C.~A., \& Eggleton, P.~P.\ 2001, \apj, 552, 664 

\bibitem[{{Nieva} \& {Przybilla}(2006)}]{nie+06}
{Nieva}, M.~F. \& {Przybilla}, N. 2006, \apjl, 639, L39

\bibitem[{{Nieva} \& {Przybilla}(2007)}]{nie+07}
---. 2007, ArXiv e-prints, 712

\bibitem[{{O'Leary} \& {Loeb}(2007)}]{ole+07}
{O'Leary}, R.~M. \& {Loeb}, A. 2007, \mnras, 1076

\bibitem[{{Paumard} {et~al.}(2006)}]{pau+06}
{Paumard}, T. {et~al.} 2006, \apj, 643, 1011

\bibitem[Perets(2009)]{per09} 
Perets, H.~B.\ 2009, \apj, 690, 795 


\bibitem[{{Perets} {et~al.}(2007){Perets}, {Hopman}, \& {Alexander}}]{per+07}
{Perets}, H.~B., {Hopman}, C., \& {Alexander}, T. 2007, \apj, 656, 709

\bibitem[{{Perets} {et~al.}(2008)}]{per+08b}
{Perets}, H.~B. {et~al.} 2008, ArXiv:0809.2087

\bibitem[Perets \& Fabrycky(2009)]{per+09} 
Perets, H.~B., \& Fabrycky, D.~C.\ 2009, arXiv:0901.4328 

\bibitem[Paczy{\'n}ski(1971)]{pac71} 
Paczy{\'n}ski, B.\ 1971, \araa, 9, 183 

\bibitem[{{Portegies Zwart}(2000)}]{por00}
{Portegies Zwart}, S.~F. 2000, \apj, 544, 437

\bibitem[{{Portegies Zwart} \& {Verbunt}(1996)}]{por+96}
{Portegies Zwart}, S.~F. \& {Verbunt}, F. 1996, \aap, 309, 179

\bibitem[{{Portegies Zwart} \& {Yungelson}(1998)}]{por+98}
{Portegies Zwart}, S.~F. \& {Yungelson}, L.~R. 1998, \aap, 332, 173

\bibitem[{{Poveda} {et~al.}(1967){Poveda}, {Ruiz}, \& {Allen}}]{pov+67}
{Poveda}, A., {Ruiz}, J., \& {Allen}, C. 1967, Boletin de los Observatorios
  Tonantzintla y Tacubaya, 4, 86

\bibitem[{{Pribulla} \& {Rucinski}(2006)}]{pri+06a}
{Pribulla}, T. \& {Rucinski}, S.~M. 2006, \aj, 131, 2986

\bibitem[{{Przybilla} {et~al.}(2008)}]{prz+08}
{Przybilla}, N. {et~al.} 2008, ArXiv: 0801.4456, \aap, in press, 801

\bibitem[{{Ramspeck} {et~al.}(2001){Ramspeck}, {Heber}, \& {Moehler}}]{ram+01}
{Ramspeck}, M., {Heber}, U., \& {Moehler}, S. 2001, \aap, 378, 907

\bibitem[{{Rolleston} {et~al.}(1999){Rolleston}, {Hambly}, {Keenan}, {Dufton},
  \& {Saffer}}]{rol+99}
{Rolleston}, W.~R.~J., {Hambly}, N.~C., {Keenan}, F.~P., {Dufton}, P.~L., \&
  {Saffer}, R.~A. 1999, \aap, 347, 69

\bibitem[Sarna \& de Greve(1996)]{sar+96} 
Sarna, M.~J., \& de Greve, J.-P.\ 1996, \qjras, 37, 11 

\bibitem[{{Savage} \& {de Boer}(1981)}]{sav+81}
{Savage}, B.~D. \& {de Boer}, K.~S. 1981, \apj, 243, 460

\bibitem[{{Schaller} {et~al.}(1992){Schaller}, {Schaerer}, {Meynet}, \&
  {Maeder}}]{sch+92a}
{Schaller}, G., {Schaerer}, D., {Meynet}, G., \& {Maeder}, A. 1992, \aaps, 96,
  269

\bibitem[{{Sembach} \& {Danks}(1994)}]{sem+94}
{Sembach}, K.~R. \& {Danks}, A.~C. 1994, \aap, 289, 539

\bibitem[{{Sesana} {et~al.}(2007){Sesana}, {Haardt}, \& {Madau}}]{ses+07b}
{Sesana}, A., {Haardt}, F., \& {Madau}, P. 2007, ArXiv:0710.4301

\bibitem[{{Sesana} {et~al.}(2008){Sesana}, {Madau}, \& {Haardt}}]{ses+08}
{Sesana}, A., {Madau}, P., \& {Haardt}, F. 2008, ArXiv e-prints

\bibitem[{{Stone}(1991)}]{sto91}
{Stone}, R.~C. 1991, \aj, 102, 333

\bibitem[{{Tokovinin} {et~al.}(2006)}]{tok+06}
{Tokovinin}, A. {et~al.} 2006, \aap, 450, 681

\bibitem[{{Tout} {et~al.}(1996)}]{tou+96}
{Tout}, C.~A. {et~al.} 1996, \mnras, 281, 257

\bibitem[{{Vanbeveren} {et~al.}(1998){Vanbeveren}, {De Loore}, \& {Van
  Rensbergen}}]{van+98b}
{Vanbeveren}, D., {De Loore}, C., \& {Van Rensbergen}, W. 1998, \aapr, 9, 63

\bibitem[{{Vrancken} {et~al.}(2000){Vrancken}, {Lennon}, {Dufton}, \&
  {Lambert}}]{vra+00}
{Vrancken}, M., {Lennon}, D.~J., {Dufton}, P.~L., \& {Lambert}, D.~L. 2000,
  \aap, 358, 639

\bibitem[{{Yu} \& {Tremaine}(2003)}]{yuq+03}
{Yu}, Q. \& {Tremaine}, S. 2003, \apj, 599, 1129

\bibitem[{Zinnecker(2005)}]{zin05}
Zinnecker, H. 2005, in Multiple Stars across the H-R Diagram, ed. S.~Hubrig,
  M.~Petr-Gotzens, \& A.~Tokovinin, 265--280

\end{thebibliography}
 
\end{document}